\documentclass[aps,10pt,prl,twocolumn,showpacs,groupedaddress]{revtex4-1}  
\usepackage{amsmath}
\usepackage{graphicx}
\usepackage{subfigure}
\usepackage[dvips]{color}

\begin{document}

\newcommand{\be}{\begin{equation}}
\newcommand{\ee}{\end{equation}}
\newcommand{\bea}{\begin{eqnarray}}
\newcommand{\eea}{\end{eqnarray}}
\newcommand{\mn}[1]{\langle #1\rangle}
\newcommand{\ket}[1]{|#1\rangle}
\newcommand{\bra}[1]{\langle #1|}
\newcommand{\pr}{\partial}

\title{Coherence of an Interacting Bose Gas: from a Single to a Double
Well}

\author{Y. Japha$^{1}$ and Y. B. Band$^{2}$}
\affiliation{$^{1}$Department of Physics, Ben-Gurion University, 
Beer-Sheva 84105, Israel\\
$^{2}$Departments of Chemistry and Electro-Optics, and the Ilse Katz
Center for Nano-Science, \\
Ben-Gurion University, Beer-Sheva 84105, Israel}
\date{\today}

\begin{abstract}
The low energy properties of a trapped bose gas split by a potential
barrier are determined over the whole range of barrier heights.  We
derive a self-consistent two-mode model which reduces, for large $N$,
to a Bogoliubov model for low barriers and to a Josephson model for
any (asymmetric) double well potential, with explicitly calculated
tunneling and pair interaction parameters.  We compare the numerical
results to analytical results that precisely specify the role of
number squeezing and finite temperatures in the loss of coherence.
\end{abstract}


\maketitle

A Bose-Einstein condensate (BEC) of ultracold atoms in a double well
potential is a model system for the study of matter wave coherence for
atom interferometry \cite{Andrews_97,Shin_04,Schumm_05,JoKetterle_07},
and tunneling and entanglement \cite{Folling_07,Esteve_08} in
many-body systems.  Its equivalence with a Josephson junction of two
superconductors separated by a tunneling barrier was demonstrated
experimentally and theoretically \cite{Albiez_05,Gati_07,Levy_07}.  As
long as the two parts of the bose gas are well connected and the
temperature is very low, most of
the atoms occupy a single spatial mode 
with a well-defined phase, satisfying the Gross-Pitaevskii equation
(GPE) \cite{GP}, while excitations and higher temperature effects are
described by Bogoliubov theory \cite{Salasnich_99, Hutchinson_00}.
However, when the tunneling rate between the two wells is low, the
system must be described in terms of two modes, e.g., within the
Bose-Hubbard model or the Josephson model, giving rise to incoherent
effects such as squeezing, phase diffusion and entanglement
\cite{Esteve_08, Gati_07, Zapata_98, Pitaevskii_01, Boukobza_08}.  But
the standard two-mode model \cite{Gati_07, Ferrini_08} allows a full
calculation of the spatial shape and the tunneling rate between the
modes only when the coupling between the two wells is small and the
potential is symmetric under inversion.

In order to gain a better qualitative and quantitative understanding
of the transition from a coherent BEC in a single well into a bose gas
separated into two parts with no relative phase relation, and study
the effects of finite temperature, we present a unified two-mode
theory which is valid over the whole range of potential barrier
heights.  We show how this theory reduces to a two-mode Bogoliubov
theory in the case of strong coupling between wells.  Moreover, if a
left-right mode representation is chosen, this unified theory reduces
to a generalized Josephson model that allows an intuitive
understanding of the coherence and number-phase uncertainties over the
whole range of system parameters.

Our calculational method is similar to recently developed theories
which can be used to calculate the statics and dynamics of the quantum
state of a bose gas in a double well \cite{Masiello_05,
Alon_Cederbaum_08}, however, the connection we make here between this
calculational method and some well-known models allows analytical
expressions for important physical properties and provides a better
understanding of the underlying physics at zero and finite
temperatures.  It lays the groundwork for a theory that allows a full
calculation of the properties of a bose gas in disconnected potentials
(e.g., optical lattices) with any number of particles.

We start from the many-particle Hamiltonian of bosons with a contact
interaction in an external potential $V({\bf r})$, $\hat{H} = \int
d^3{\bf r} \, \hat{\psi}^{\dag}({\bf r}) \left[H_0({\bf r}) +
\frac{1}{2} g \hat{\psi}^{\dag}({\bf r}) \hat{\psi}({\bf r}) \right]
\hat{\psi}({\bf r})$, where $H_0 = p^2/2m+V$ is the single-particle
Hamiltonian, $\hat{\psi}$ is the field operator, $g = 4\pi \hbar^2
a/m$ is the interaction constant, $a$ is the $s$-wave scattering
length and $m$ is the particle mass.  We expand the field operator in
terms of a finite set of orthogonal spatial modes $\phi_j({\bf r})$,
$\hat{\psi}=\sum_j \phi_j \hat{a}_j$, with $\hat{a}_j$ being bosonic
annihilation operators.
The Hamiltonian can then be written as
\begin{equation}  \label{eq:Ham1}
    \hat{H} = \sum_{i,j}\epsilon_{ij}\hat{a}_i^{\dag}\hat{a}_j +
    \frac{1}{2} \sum_{ijkl}U_{ijkl}\hat{a}_i^{\dag}\hat{a}_j^{\dag}
    \hat{a}_k\hat{a}_l ,
\end{equation}
with $\epsilon_{ij} = \mn{\phi_i|\hat{H}_0|\phi_j}$ and
$U_{ijkl}=g\int d^3{\bf r}\ \phi^*_i \phi^*_j \phi_k\phi_l$.
Equations of motion for the spatial modes $\phi_i$ may be obtained by
multiplying the Heisenberg equations of motion for the field operator,
$i\hbar\partial \hat{\psi}/\partial t=[\hat{H},\hat{\psi}]$, by
$\hat{a}_i^{\dag}$ from the left and taking the expectation value.
Then, using $i\hbar\dot{\hat{a}}_i=[\hat{H},\hat{a}_i]$ we obtain
\begin{equation}  \label{eq:modes} 
    i\hbar\frac{\partial \phi_i}{\partial t}= {\cal P} \left[H_0\phi_i
    + g\sum_{jklm} (\rho^{-1})_{im} \rho_{mklj} \phi_k^* \phi_l\phi_j
    \right] .
\end{equation}
Here ${\cal P}=1-\sum_n\ket{\phi_n}\bra{\phi_n}$ is an operator
projecting onto the function subspace which is not spanned by the
finite set $\{\phi_j\}$ and $\rho_{ij} =
\mn{\hat{a}_i^{\dag}\hat{a}_j}$ and $\rho_{mklj} =
\mn{\hat{a}_m^{\dag} \hat{a}_k^{\dag} \hat{a}_l\hat{a}_j}$ are the
single-particle and the two-particle reduced density matrices, which
should be determined self-consistently with the solution for the mode
functions.  Equation~(\ref{eq:modes}), which was previously obtained
using a variational principle \cite{Masiello_05, Alon_Cederbaum_08} is
automatically satisfied if $\{\phi_j\}$ are a complete set, since then
${\cal P} = 0$; if only a single mode is considered, (\ref{eq:modes})
reduces to the GPE, which is suitable for a many boson system well
below the condensation temperature in a single well.  Here we consider
the steady state solutions, where $\partial \phi_i/\partial t = 0$,
for the case of two modes and show how it leads to a Bogoliubov theory
in one limit and to the Josephson model otherwise.

First, note that the Heisenberg equations for $\phi_j$ and $\hat{a}_j$
following from Eqs.~(\ref{eq:Ham1}) and~(\ref{eq:modes}) are invariant
under the unitary transformation $\phi_j\to \sum_k {\cal
U}_{jk}\phi_k$, $\hat{a}_j\to \sum_k{\cal U}^*_{kj}\hat{a}_k$.  With
two modes, we may choose the basis $\phi_L$ and $\phi_R$ with minimum
overlap $\int d^3{\bf r}|\phi_L({\bf r})||\phi_R({\bf r})|$, that
mainly occupy the left or the right well, respectively.
Alternatively, we may choose a basis $\phi_c = \cos(\theta/2)\phi_L+
\sin(\theta/2)\phi_R$ and $\phi_e=-\sin(\theta/2)\phi_L+\cos(\theta/2)
\phi_R$ with $\theta$ chosen such that $\langle
\hat{a}_c^{\dag}\hat{a}_e\rangle =0$.  $\phi_c$ and $\phi_e$ represent
a condensate mode and an excitation mode.  When the atomic cloud is
well connected, the condensate mode (symmetric mode in the case of a
symmetric potential) is dominantly occupied.  We then define the
operator $\hat{b}\equiv \hat{a}_c^{\dag} \hat{a}_e/\sqrt{\langle
n_c\rangle}$, which transfers a particle from the excited mode to the
condensate mode, where $\langle \hat{n}_c\rangle$ is the number of
condensate particles.  When $\langle \hat{n}_c\rangle \gg \langle
\hat{n}_e\rangle$, $\hat{b}\approx \hat{a}_e$ approximately satisfies
bosonic commutation relations $[\hat{b},\hat{b}^{\dag}]\approx 1$.
Using this approximation, the two-mode form in Eq.~(\ref{eq:Ham1})
reduces to the Bogoliubov form,
\begin{equation} \label{eq:HamBog}
    \hat{H}_{\rm B} = A\hat{b}^{\dag}\hat{b}+\frac{1}{2}
    B(\hat{b}\hat{b}+\hat{b}^{\dag}\hat{b}^{\dag}) .
\end{equation}
Under the same conditions, Eq.~(\ref{eq:modes}) for $\phi_c$ becomes
the GPE, and in steady-state, for $\phi_e$ we obtain
\begin{equation} 
    [\hat{H}_0+2g\langle \hat{n}_c\rangle \frac{\langle
    \hat{b}^{\dag}\hat{b} \rangle}{\langle \hat{n}_e\rangle}
    |\phi_c|^2-E_e]\phi_e +g\langle \hat{n}_c\rangle \frac{\langle
    \hat{b}\hat{b}\rangle^*}{\langle \hat{n}_e\rangle}
    \phi_c^2\phi_e^* = 0.
\end{equation}
The steady state values of $\langle \hat{b}^{\dag}\hat{b}\rangle$ and
$\langle \hat{b}\hat{b}\rangle$ are obtained by diagonalizing
Hamiltonian~(\ref{eq:HamBog}) using the Bogoliubov transformation
$\hat{b}=u\hat{\alpha}-v^* \hat{\alpha}^{\dag}$, where $\hat{\alpha}$
is a bosonic quasi-particle annihilation operator and $u$,$v$ satisfy
$|u|^2-|v|^2=1$.  This yields the excitation energy $E_{\rm
ex}=\sqrt{A^2-B^2}$ and the ground state expectation values $\langle
\hat{n}_e\rangle = \langle \hat{b}^{\dag} \hat{b}\rangle = |v|^2$ and
$\langle \hat{b} \hat{b}\rangle = -uv^*$, with
$|v|^2=\frac{1}{2}\left(A/E_{\rm ex}-1\right)$.  Below we show the
equivalence of this two-mode form of the Bogoliubov theory to the
Josephson model, and discuss its relation with the conventional
Bogoliubov theory with two quasi-particle functions $u({\bf r})$ and
$v({\bf r})$ \cite{Salasnich_99, Hutchinson_00}.

Let us now return to the left-right representation.  In
Ref.~\cite{Gati_07}, the Josephson Hamiltonian was derived from
Hamiltonian~(\ref{eq:Ham1}) in its two-mode form for a symmetric
double well potential.  However, the equations of motion for the mode
functions, which yield non-orthogonal modes, are valid only in the
limit of weak coupling between the wells.  Here we present an
alternative derivation valid in the general case and compare the
results to the full two-mode theory that follows from
Eqs.~(\ref{eq:Ham1}) and (\ref{eq:modes}).

Consider a system of $N$ particles which can occupy either mode
$\phi_L$ or $\phi_R$.  We define $\hat{n} \equiv \hat{a}_L^{\dag}
\hat{a}_L$ as the number operator for particles in the left mode
$\phi_L$, with eigenstates $\ket{n}$ having $n$ particles in the left
mode and $N-n$ in the right mode.  A general state of the system may
be written as $\sum_n c_n\ket{n}$, with $\sum_n |c_n|^2=1$.  We define
the phase operator as $e^{i\hat{\varphi}}\equiv \sum_{n=0}^{N-1}
\ket{n+1}\bra{n}+\ket{0}\bra{N}$, that has eigenstates
$\ket{\varphi}=\sum_n e^{-in\varphi}\ket{n}$ and eigenvalues
$\varphi_k = 2\pi k/N$ with integer $k$.  It can be shown that when
the extreme states $\ket{0}$ and $\ket{N}$ in which all the particles
occupy either the left or right modes are not significantly populated,
the phase and the number operators satisfy conjugate commutation
relations, $[\hat{\varphi}, \hat{n}] = i$, similar to position and
momentum.  When $N$ and $n$ are large, $a_L^{\dag}a_R \approx
\sqrt{\hat{n}(N-\hat{n})} \, e^{i\hat{\varphi}}$, and, after
neglecting constant terms, Hamiltonian~(\ref{eq:Ham1}) goes to,
\begin{eqnarray}  
\hat{H} &\approx & (\bar{\epsilon}_L-\bar{\epsilon}_R) \hat{n}
-U\hat{n}(N-\hat{n}) - 2J\sqrt{\hat{n}(N-\hat{n})}
\cos\hat{\hat{\varphi}}\nonumber \\
&& +2U_{LRLR}\hat{n}(N-\hat{n})(\cos\hat{\varphi}-2\langle
\cos\hat{\varphi}\rangle) \cos\hat{\varphi},
\label{eq:Hamb}
\end{eqnarray}
where $U\equiv (U_L+U_R-2U_{LRLR})/2$, $\bar{\epsilon}_i =
\epsilon_{ii}+U_i(N-1)/2$, and $U_i\equiv U_{iiii}$.  The tunneling rate
$J$ is approximately
\begin{equation}  \label{eq:J}
    J = -\int d^3 {\bf r}\phi_L^*({\bf r}) \left[\hat{H}_0 + g\langle
    \hat{n}({\bf r}\rangle)\right] \phi_R({\bf r}).
\end{equation}
Here $\langle \hat{n}({\bf r})\rangle\equiv \langle
\hat{\psi}^{\dag}({\bf r}) \hat{\psi}({\bf r})\rangle$ is the particle
density, $\hat{\psi}({\bf r})\approx \sqrt{\hat{n}}\phi_L({\bf r}) +
\sqrt{N-\hat{n}}\phi_R({\bf r}) e^{i\hat{\varphi}}$, and $g \langle n
\rangle$ is the mean-field potential, which should be
self-consistently calculated with the solution of the Hamiltonian.  We
expand the Hamiltonian~(\ref{eq:Hamb}) around the values of $n=n_0$
and $\varphi=0$ that minimize the energy $\langle \hat{H} \rangle$,
and find that the last term in (\ref{eq:Hamb}) contributes only fourth
order corrections or higher in $\hat \varphi$.  This term is also
negligible when the overlap between $\phi_L$ and $\phi_R$ is small and
$U_{LRLR} \to 0$.  One can then approximate $\hat{H}$ by the Josephson
hamiltonian
\begin{equation} \label{eq:HJosephson} 
    \hat H_J = \tilde{U} (\hat{n}-n_0)^2 + JN \eta
    (1-\cos\hat{\varphi}) ,
\end{equation} 
where $\eta \equiv 2\sqrt{n_0(N-n_0)}/N$ ($\eta=1$ for the symmetric
case where $n_0=N/2$) and $\tilde{U} = U+2J/N\eta^3$.

\begin{figure}
\includegraphics[width=0.42\textwidth]{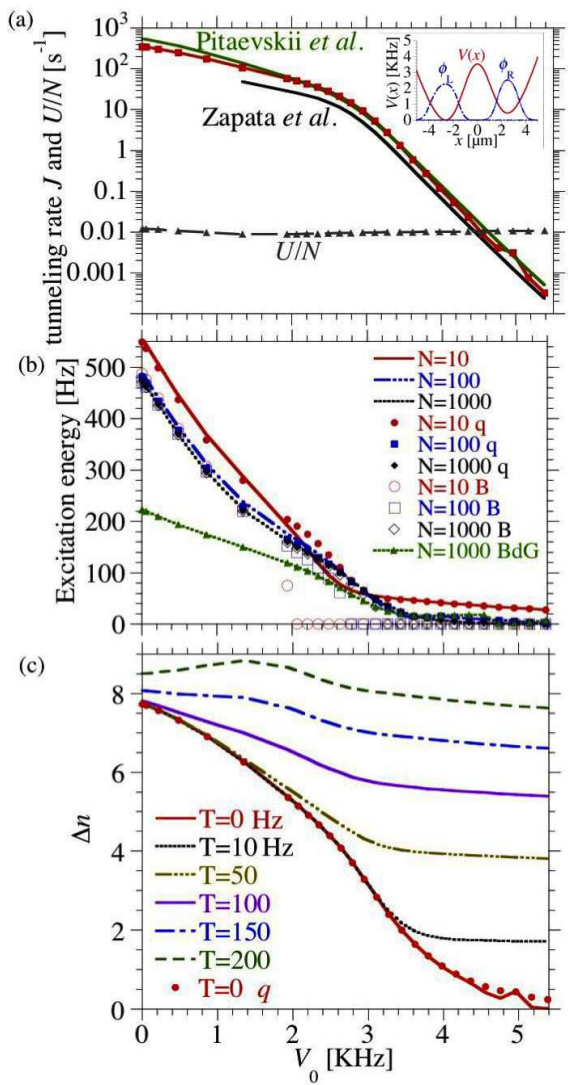}
\caption{(color online) (a) Tunneling rate $J$ and scaled pair
interaction energy $U/N$ as a function of barrier height $V_0$ for an
asymmetric 1D double-well potential with $N=1000$ atoms.  Our result
[Eq.~\protect(\ref{eq:J})] (solid red curve with squares at the
computed points) is compared to the semiclassical results of Zapata
{\it et }~\protect\cite{Zapata_98} using GPE, and the expression of
Pitaevskii et al.~\protect\cite{Pitaevskii_01} using the modes
$\phi_L$ and $\phi_R$ calculated here.  Inset: $V(x)$ and the mode
functions for a given $V_0$.  (b) Energy $E_{\mathrm{ex}}$ of the
lowest excitation and (c) number uncertainty $\Delta n$ of the bose
gas.  The numerical solutions of Eq.~(\ref{eq:Ham1}) (solid curves)
are compared to the analytic quadratic approximation to the Josephson
Hamiltonian (``q''), excitations of the two-mode Bogoiubov model
[``B'', Eq.~(\ref{eq:HamBog})] and the Bogoliubov-de Gennes
calculation (``BdG'').}
\label{fig:J}
\end{figure}

To gain a physical picture of the transition from a single well
coherent gas into a two-mode gas separated into two wells, we solved
Eqs.~(\ref{eq:Ham1}) and~(\ref{eq:modes}) for a model system reduced
to one dimension, using potential and interaction parameters of the
same order as those used in experiments reported in
Refs.~\cite{Levy_07, Reichel_10} with $^{87}$Rb atoms.  We compare
these results to analytical predictions of the Bogoliubov and
Josephson models above.  We use an asymmetric potential $V(x) =
\frac{1}{2} m\omega^2 (x-x_0)^2 + V_0 \, e^{-x^2/\sigma_b^2}$ where
the harmonic part has a frequency $\omega = 2\pi \times 224$ Hz, shift
$x_0=-0.2$, and Gaussian barrier width $\sqrt{2} \sigma_b = 5 \, \mu$m
with a varying peak energy $V_0$.  For the atom-atom interaction we
use $g_{1D}N= 2.36 \times 10^{-36}$ J m, such that the mean-field
interaction potential (and hence the form of the spatial modes) is
similar for varying particle numbers.  In Fig.~\ref{fig:J}(a) we show
the tunneling rate $J$ (for $N=1000$) calculated from
Eq.~(\ref{eq:J}) as a function of $V_0$.  We compare this result with
a semiclassical calculation based on Eq.~(9) of Ref.~\cite{Zapata_98}
with the atomic density $\rho$ from a GPE solution, and with the
expression $J=(\hbar^2/m)\left[\phi_L\partial \phi_R/\partial
x-\phi_R\partial \phi_L/\partial x \right]_{x=0}$
\cite{Pitaevskii_01}.  The parameter $J$ drops exponentially with
barrier height over the whole tunneling regime ($V_0>3$KHz), while the
pair interaction parameter $U$ is approximately constant.

The ground state properties following from the
Hamiltonian~(\ref{eq:HJosephson}) are determined by the relative
magnitude of the pair interaction parameter $U$, which tends to reduce
number uncertainty and the tunneling term, which tends to reduce phase
uncertainty and hence increase number uncertainty.  In the region
where $JN\eta>U$ [to the left of the intersection between $J$ and
$U/N$ in Fig.~\ref{fig:J}(a)] the number uncertainty, $\Delta
n^2\equiv \sum_n |c_n|^2(n-n_0)^2$, is wide enough so that
$\Delta\varphi\ll 1$ and the $1-\cos\hat{\varphi}$ may be approximated
by $\hat{\varphi}^2/2$ in (\ref{eq:HJosephson}).  The Hamiltonian then
has a harmonic oscillator form, and the ground state and lowest
excitations may be approximated by
\begin{equation} \label{eq:Ckn}
    c^{(k)}_n = \frac{H_k\left(\frac{n-n_0}{\sigma}\right)}
    {(\pi\sigma^2)^{1/4}}
    e^{-\left[\frac{(n-n_0)^2}{2\sigma^2}\right]} , \, \, \sigma =
    \left(\frac{JN\eta}{2\tilde{U}}\right)^{1/4},
\end{equation}
where $H_k(x)$ is the $k$th order Hermite polynomial and
$\sigma/\sqrt{2}$ is the number uncertainty $\Delta n$ of the ground
state.  These eigenstates have energies
$E_k=E_{\mathrm{ex}}(k+\frac{1}{2})$ with $E_{\mathrm{ex}} = (2JN
\eta\tilde{U})^{1/2}$, number variance $\langle \Delta n^2\rangle_k=
\sigma^2(k+\frac{1}{2})$ and phase variance $\mn{\Delta\varphi^2}_k =
\sigma^{-2}(k+\frac{1}{2})$.  In the weak interaction limit, $U \ll
J/N$, Eq.~(\ref{eq:Ckn}) approximates a coherent state with $c_n =
\binom{N}{n}$$^{1/2} \sin^n(\theta/2) \cos^{N-n}(\theta/2)$ and
$\eta=\sin\theta$.  Number squeezing is characterized by the ratio
$\xi \equiv \Delta n/\Delta n_{U=0} = (2J/N\tilde{U}\eta^3)^{1/4}$
between the number uncertainties of a general ground state
in~(\ref{eq:Ckn}) and the Poissonian width $\Delta
n_{U=0}=\sqrt{N}\eta/2$ of the coherent state.

In the strong tunneling regime, transformation of
Hamiltonian~(\ref{eq:Hamb}) into the condensate-excitation
representation using a rotation angle $\theta \approx {\rm
arcsin}\eta$, neglecting terms of order 3 or higher in $\hat{a}_e$ and
$\hat{a}_e^{\dag}$, yields the Bogoliubov Hamiltonian
(\ref{eq:HamBog}) with $A = \frac{1}{2}UN\eta^2+2J/\eta$ and
$B=\frac{1}{2}UN\eta^2$.  Hence, the magnitude of the quasi-particle
factor $v$ represents the amount of number squeezing, with
$v=\frac{1}{2}(1/\xi-\xi)$.

The properties of the system are illustrated in Figs.~\ref{fig:J}(b)
and (c) for different particle numbers and temperatures.  As the
barrier height $V_0$ grows, number squeezing becomes stronger and the
excitation energy $E_{\rm ex}$ drops to the point where $J<U/N$.  In
this ``Fock regime'', $\Delta n< 1$ and $E_{\mathrm{ex}}\to
U|2(n_{0+}-n_0)-1|$, where $n_{0+}$ is the closest integer greater
than or equal to $n_0$ and the quadratic approximation
in~(\ref{eq:HJosephson}) breaks down.  Good agreement is found between
excitation energies obtained from the full calculation in the strong
tuneling regime to the analytic results of the Josephson and
Bogoliubov models.  However, solutions of the Bogoliubov-de Gennes
equations \cite{Salasnich_99} which are not restricted to one
excitation mode yield lower excitation energies.  As the barrier
height grows and $E_{\rm ex}$ drops, the single real-particle
excitation mode $\phi_e$ becomes dominant, hence the quasi-particle
functions $u({\bf r})$ and $v({\bf r})$ become more similar to
$\phi_e({\bf r})$ and the two-mode predictions are more accurate.

\begin{figure}
\includegraphics[width=0.42\textwidth]{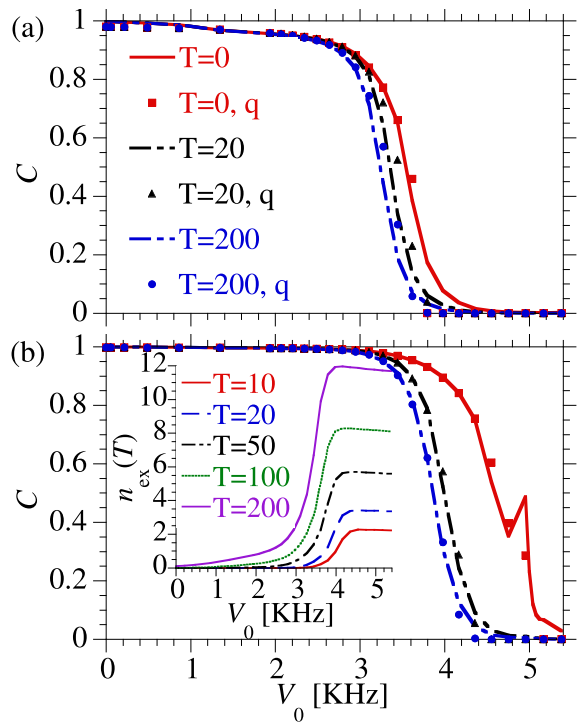}
\caption{Coherence $C$ at several temperatures as a function of the
barrier height for (a) $N=100$ atoms, and (b) $N=1000$ atoms.  Solid
curves: coherence function $g^{(1)}(x_L,x_R)$, with $x_L$ and $x_R$ in
the left and right wells respectively.  Symbols: quadratic
approximation results [Eq.~(\protect\ref{eq:C_T})].  The inset in (b)
shows the number of excitations $n_{\mathrm{ex}}(T)$ vs.~$V_0$ for
different $T$.}
\label{fig:coherence}
\end{figure}

The coherence of a split bose gas, given by the correlation function
$g^{(1)}({\bf r},{\bf r}') = \mn{\hat{\psi}^{\dag}({\bf r})\hat{\psi}({\bf
r}')}/\sqrt{n({\bf r})n({\bf r}')}$ with ${\bf r}$ and ${\bf r}'$ on
opposite sides of the barrier, determines the repeatability of
interference fringes when the potential is dropped and the two parts
begin to overlap.  If the modes $\phi_L$,$\phi_R$ are well separated,
it may be approximated by $C\equiv \langle
\hat{a}_L^{\dag}\hat{a}_R\rangle/\sqrt{n_Ln_R}\approx \langle
e^{i\hat{\varphi}}\rangle$ \cite{Pitaevskii_01}.  At finite
temperature, the excitation modes of Eq.~(\ref{eq:Ckn}) are populated
according to the Bolzmann distribution, $P_k\propto e^{-E_k/k_BT}$,
and the quadratic approximation to the Josephson Hamiltonian yields a
Gaussian phase distribution.  Hence the coherence may be approximated
by \cite{Stern_90}
\begin{equation}   \label{eq:C_T}
    C\approx \mn{e^{i\hat\varphi}}_T \approx e^{-\frac{1}{2}\langle \Delta
    \hat{\varphi}^2\rangle} = e^{-\frac{1}{2\sigma^2}
    \left[n_{\mathrm{ex}}(T) + 1/2 \right]},
\end{equation}
where $n_{\rm ex}(T) = (e^{-E_{\rm ex}/k_B T}-1)^{-1}$ ($\neq n_e$) is
the mean number of excitations.  The term proportional to $n_{\rm ex}$
in the exponent of (\ref{eq:C_T}) is the finite temperature ($n_{\rm
ex}>0$) contribution to the dephasing, while the second term
$1/4\sigma^2$ is the contribution of ground-state number squeezing to
dephasing.  In the regime where the Bogoliubov model is suitable, loss
of coherence is equivalent to the population of real particle
excitations, as $C \approx 1-2n_e/N\eta^2$, where $n_e\approx \langle
\hat{b}^{\dag}\hat{b}\rangle = |v|^2+n_{\rm ex}(|u|^2+|v|^2)$ again
includes the effect of ground state squeezing and thermal population
of excitations.

Figure~\ref{fig:coherence} compares the analytical
expression~(\ref{eq:C_T}) for the coherence to the full calculation of
$g^{(1)}(x,x')$ with $x$ and $x'$ in the middle of the left and right
wells, respectively.  It demonstrates that the analytical expressions
are valid for low temperatures and large $N$.  The cusp in
Fig.~\ref{fig:coherence}(b) [and \ref{fig:J}(c)] results when $n_0$ is
close to a half integer.

In summary, we developed a two-mode theory for a bose gas in a double
well that is well-suited to all barrier heights.  At low barrier
heights and relatively weak interactions, only one mode is
macroscopically occupied, hence results using the GPE agree with our
two-mode theory.  Comparison of the full numerical results to
analytical solutions for the Josephson model with explicitly
calculated parameters, shows good agreement for large $N$.  Comparison
of the excitation energies computed with our theory and with a
Bogoliubov-de Gennes model, which is not restricted to two-modes,
shows good agreement at medium barrier heights, but only qualitative
agreement for low barriers.  Our approach will enable development of
models for calculating and understanding the steady-state and the
time-dependent properties of many-body systems with disconnected
potentials and arbitrarily large number of particles.


This work was supported by grants from the U.S.-Israel Binational
Science Foundation (No.~2006212), the Israel Science Foundation
(No.~29/07), and the James Franck German-Israel Binational Program.

\end{document}